\documentclass[sigconf]{acmart}

\usepackage{booktabs} 
\usepackage{lipsum}
\usepackage{multirow}
\usepackage{array}
\usepackage{courier}
\usepackage{mathtools}
\usepackage{subcaption}
\usepackage{amsfonts}
\usepackage{tikz}
\usepackage{mathtools}
\DeclarePairedDelimiter{\abs}{\lvert}{\rvert}
\DeclarePairedDelimiter\norm{\lVert}{\rVert}%

\copyrightyear{2017} 
\acmYear{2017} 
\setcopyright{acmcopyright}
\acmConference{SIGIR '17}{}{August 07-11, 2017, Shinjuku, Tokyo, Japan}\acmPrice{15.00}\acmDOI{http://dx.doi.org/10.1145/3077136.3080790}\acmISBN{978-1-4503-5022-8/17/08}

\begin{document}
\title{Learning to Rank Question Answer Pairs with Holographic Dual LSTM Architecture} 

\author{Yi Tay}
\affiliation{%
  \institution{Nanyang Technological University}
}
\email{ytay017@e.ntu.edu.sg}

\author{Minh C. Phan}
\affiliation{
  \institution{Nanyang Technological University}
}
\email{phan0005@e.ntu.edu.sg}

\author{Luu Anh Tuan}
\affiliation{%
 \institution{Institute for Infocomm Research}
}
\email{at.luu@i2r.a-star.edu.sg}

\author{Siu Cheung Hui}
\affiliation{%
  \institution{Nanyang Technological University}
}
\email{asschui@ntu.edu.sg}

\begin{abstract}
We describe a new deep learning architecture for learning to rank question answer pairs. Our approach extends the long short-term memory (LSTM) network with holographic composition to model the relationship between question and answer representations. As opposed to the neural tensor layer that has been adopted recently, the holographic composition provides the benefits of scalable and rich representational learning approach without incurring huge parameter costs. Overall, we present Holographic Dual LSTM (HD-LSTM), a unified architecture for both deep sentence modeling and semantic matching. Essentially, our model is trained end-to-end whereby the parameters of the LSTM are optimized in a way that best explains the correlation between question and answer representations. In addition, our proposed deep learning architecture requires no extensive feature engineering. Via extensive experiments, we show that HD-LSTM outperforms many other neural architectures on two popular benchmark QA datasets. Empirical studies confirm the effectiveness of holographic composition over the neural tensor layer.
\end{abstract}

\renewcommand{\footnotesize}{\scriptsize}


\maketitle

\section{Introduction}

Learning to rank techniques are central to many web-based question answering (QA) systems such as factoid-based QA or community-based QA (CQA). In these applications, questions are matched against an extensive database to find the most relevant answer. Essentially, this is highly related to many search and information retrieval tasks such as traditional document retrieval and text matching. However, a key difference is that questions and answers are often much shorter compared to full-fledged documents whereby the problem of lexical chasm \cite{DBLP:conf/sigir/BergerCCFM00,DBLP:conf/sigir/XueJC08,DBLP:conf/cikm/JeonCL05} becomes more prevalent. As such, this makes the already difficult task of designing features for questions and answers even harder. Furthermore, traditional approaches often involve extensive handcrafted features and domain expertise which can be laborious and expensive. In addition, constructing features from textual clues \cite{DBLP:conf/sigir/WangMC09,DBLP:conf/acl/ZhouCZL11,DBLP:journals/coling/SurdeanuCZ11,DBLP:conf/naacl/YaoDCC13} such as lexical and syntactic features is difficult and provide limited benefits. Overall, the challenges of learn-to-rank QA systems are two-fold. First, feature representations of questions and answers have to be learned or designed. Second, a similarity function has to be defined to match questions to answers. 

 Recently, deep learning architectures have been an extremely popular choice for learning distributed representations of words, sentences or documents \cite{DBLP:conf/nips/MikolovSCCD13,DBLP:journals/taslp/PalangiDSGHCSW16}. Generally, this is known as representational learning whereby low dimensional vectors are learned for words or documents via neural networks such as the convolutional neural networks (CNN), recurrent neural network (RNN) or standard feed-forward multi-layer perceptron (MLP). This has widespread applications in the field of NLP and IR such as semantic text matching \cite{DBLP:conf/aaai/WanLGXPC16,DBLP:conf/sigir/SeverynM15}, relation detection \cite{DBLP:conf/emnlp/LuuTHN16}, language modeling \cite{DBLP:journals/taslp/PalangiDSGHCSW16} and question answering \cite{DBLP:conf/sigir/SeverynM15,DBLP:conf/ijcai/QiuH15}. Essentially, the attractiveness of these models stem from the fact that features are learned in deep learning architectures in an end-to-end fashion and often require little or no human involvement. Furthermore, the performance of these models are often spectacular. 

Additionally and recently, it has also been fashionable to model the relationship between vectors via tensor layers \cite{DBLP:conf/sigir/XiaXLGC16,DBLP:conf/aaai/WanLGXPC16,DBLP:conf/ijcai/QiuH15}. A recent work, the convolutional neural tensor network (CNTN) \cite{DBLP:conf/ijcai/QiuH15} demonstrates impressive results on community-based question answering. In their work, a convolutional neural network is used to learn representations for questions and answers while a tensor layer is used to model the relationship between the representations using an additional tensor parameter. This is powerful because the tensor layer models multiple views of dyadic interactions between question and answer pairs which enables rich representational learning. Overall, the CNTN is a unified architecture that learns representations and performs matching in an end-to-end fashion.

However, the use of a tensor layer may not be implication free. Firstly, adding a tensor layer severely increases the number of parameters which naturally and inevitably increases the risk of overfitting. Secondly, this significantly increases computational and memory cost of the overall network. Thirdly, the inclusion of a tensor layer also \textit{indirectly} restricts the expressiveness of the QA representations since increasing the parameters of the QA representations would easily incur memory and computational costs of quadratic scale at the tensor layer.

 In lieu of the above mentioned weaknesses, we propose an alternative to the tensor layer. For the first time, we adopt holographic composition to model the relationship between question and answer embeddings. Our approach is largely based on holographic models of associative memory and employs circular correlation to learn the relationship between QA pairs. The prime contributions of our paper can be summarized as follows:

\begin{itemize}

\item For the first time, we adopt holographic composition for modeling the interaction between representations of QA pairs. Unlike the tensor layer, our compositional approach is essentially parameterless, memory efficient and scalable. Furthermore, our approach also enables rich representational learning by employing circular correlation. 

\item As a whole, we present a novel deep learning architecture, HD-LSTM (Holographic Dual LSTM) for learning to rank QA pairs. Our model is a unified architecture similar to \cite{DBLP:conf/ijcai/QiuH15}. However, instead of the CNN, we use multi-layered long short-term memory neural networks to learn representations for questions and answers. Similar to other deep learning models, our approach does not require extensive feature engineering or domain knowledge. 

\item We provide extensive experimental evidence of the effectiveness of our model on both factoid question answering and community-based question answering. Our proposed approach outperforms many other neural architectures on TREC QA task and on the Yahoo CQA dataset. 
\end{itemize}

\section{Related Work}

Our work is concerned with ranking question and answer pairs to select the most suitable answer for each question. Across the rich history of IR research, techniques for doing so have been primarily focused on lexical and syntactic feature-dependent approaches. These techniques include the Tree Edit Distance (TED) model \cite{DBLP:conf/naacl/HeilmanS10a}, Support Vector Machines (SVMs) with tree kernels \cite{DBLP:conf/sigir/SeverynMTBR14} and linear chain Conditional Random Fields (CRFs) \cite{DBLP:conf/naacl/YaoDCC13} with features from the TED model. However, apart from relying heavily on handcrafted features such as cumbersome parse trees, these approaches have limited performance and have been shown to be outclassed by modern deep learning approaches such as convolutional neural networks \cite{DBLP:conf/sigir/SeverynM15,DBLP:journals/corr/YuHBP14}. 

The key intuition behind deep learning architectures is to learn low-dimensional representations of words, documents or sentences which can be used as input features. For example, Yu et al. \cite{DBLP:journals/corr/YuHBP14} employed a convolutional neural network for feature learning of QA pairs and subsequently applied logistic regression for prediction. Despite its simplicity, the performance of Yu et al. has already surpassed all traditional approaches \cite{DBLP:conf/sigir/WangMC09,DBLP:conf/emnlp/WangSM07,DBLP:conf/naacl/HeilmanS10a,DBLP:conf/coling/WangM10a,DBLP:conf/sigir/SeverynMTBR14}. Another attractive quality of deep learning architectures is that features can be learned in an end-to-end fashion. Severyn et al. \cite{DBLP:conf/sigir/SeverynM15} demonstrated a unified architecture that trains a convolutional neural network together with a multi-layer perceptron. In short, features are learned while the parameters of the network are optimized for the task at hand. 

In the architectures of Severyn et al. \cite{DBLP:conf/sigir/SeverynM15} and Yu et al. \cite{DBLP:journals/corr/YuHBP14}, representations of questions and answers are learned separately and concatenated for prediction at the end. Qiu et al. \cite{DBLP:conf/ijcai/QiuH15} introduced a tensor layer to model the relationship between question and answer representations. The tensor layer can be seen as a compositional technique to learn the relationship between two vectors and was adapted from the neural tensor network (NTN) by Socher et al. \cite{DBLP:conf/nips/SocherCMN13,socher2013recursive}. The NTN was originally incepted in the field of NLP for semantic parsing and used as a compositional operator in recursive neural tensor networks (RNTN) \cite{socher2013recursive} and also relational learning on knowledge bases \cite{DBLP:conf/nips/SocherCMN13}. It has also recently seen adoption for modeling document novelty in \cite{DBLP:conf/sigir/XiaXLGC16}. The tensor layer models multiple dyadic interactions between two vectors via an additional tensor parameter. This suggests rich representational learning that is useful for matching text pairs. 

Additionally, recurrent neural networks such as the long short-term memory (LSTM) networks are also widely popular for learning sentence representations and has seen wide adoption in a variety of NLP tasks. Without an exception, LSTM networks are also widely adopted in QA \cite{DBLP:conf/aaai/MuellerT16,DBLP:conf/acl/WangN15,DBLP:conf/aaai/WanLGXPC16}. The usage of grid-wise similarity matrices within neural architectures are also recently very fashionable and have seen wide adoption\footnote{Notably, our proposed holographic composition can also be used for grid-wise matching.} in QA tasks \cite{DBLP:conf/emnlp/ParikhT0U16,DBLP:conf/aaai/WanLGXPC16,he2015multi} to model the interactions between QA pairs. For example, in the MV-LSTM \cite{DBLP:conf/aaai/WanLGXPC16}, all positional hidden states from both LSTMs are being matched grid-wise using a variety of similarity scoring functions followed by a max-pooling layer. On the other hand, the works of \cite{DBLP:journals/corr/XiongZS16} are concerned with learning grid-wise attentions. On a side note, it is good to note that, grid-wise matching, though highly competitive, naturally incurs a prohibitive computational cost of quadratic scale. 

As seen in many recent works, the tensor layer is highly popular to model relationship between two vectors \cite{DBLP:conf/ijcai/QiuH15,DBLP:conf/aaai/WanLGXPC16}. However, a tensor layer adds a significant amount of parameters to the network causing implications in terms of runtime, memory, risk of overfitting as well as an inevitable restriction of flexibility in designing representations for questions and answers. Specifically, increasing the dimensionality of the LSTM or CNN output by $x$ would incur a parameter cost of $x^2$ in the tensor layer which can be non-ideal especially in terms of scalability. As an alternative to the tensor layer, our novel deep learning architecture adopts the circular correlation of vectors to model the interactions between question and answer representations. The circular correlation of vectors, along with circular convolution, are typically used in Holography to store and retrieve information \cite{Gabor:1969:AHM:1663617.1663619,DBLP:journals/tnn/Plate95} and are also known as correlation-convolution (holographic-like) memories. Due to its connections with holographic models of associative memories \cite{DBLP:journals/tnn/Plate95}, we refer to our model as Holographic Dual LSTM. It is good to note that a similar but fundamentally different work \cite{DBLP:conf/nips/Plate92} also used holography inspired operations \textit{within} recurrent neural networks. However, our work is the first work to incorporate holographic representational learning for QA embeddings. 

In addition, holographic composition \cite{DBLP:journals/tnn/Plate95} can also be interpreted as compressed tensor product which also enables rich representational learning \textbf{without} severely increasing the number of parameters of the network. In this case, the parameters of the network are learned in a way that best explains the correlation between questions and answers. In the same domain where the neural tensor network was incepted, holographic embeddings of knowledge graphs \cite{DBLP:conf/aaai/NickelRP16}, demonstrates the effectiveness of holographic composition in the task of relational learning on knowledge bases. As a whole, we propose a novel deep learning architecture based on long short-term memory neural networks while using holographic composition to model the interactions between QA embeddings, this enables rich representational learning with improved flexibility and scalability. The outcome is a highly performant end-to-end deep learning architecture for learning to rank for QA applications. 

\section{Preliminaries}
In this section, we introduce the background for the remainder of the paper. Namely, we formally give the problem definition and introduce fundamental deep learning models required to understand the remainder of the paper.

\subsection{Problem Statement and Approach}
The task of supervised learning to rank can be typically regarded as a binary classification problem. Given a set of questions $q_i \in Q$, the task is to rank a list of candidate answers $a_i \in A$. Specifically, we try to learn a function $f(q,a)$ that outputs a relevancy score $f(q,a) \in [0,1]$ for each question answer pair. This score is then used to rank a list of possible candidates. Typically, there are three different ways for supervised text ranking, namely, pairwise, pointwise and listwise. Pairwise considers maximizing the margin between positive and negative question-answer pairs with an objective function such as the hinge loss. Pointwise considers each pair, positive or negative, individually. On the other hand, listwise considers a question and all candidate answers as a training instance. Naturally, pairwise and listwise are much harder to train, implement and take a longer time due to having to process more instances. Therefore, in this work, we mainly consider a pointwise approach when designing our deep learning model.

\subsection{Long short-term Memory (LSTM)}
First, we introduce the Long Short-Term Memory (LSTM) \cite{hochreiter1997long}. LSTMs are a type of recurrent neural network that are capable of learning long term dependencies across sequences. Given an input sentence $S=(x_o,x_1...,x_n)$, the LSTM returns a sentence embedding $h_t$ for position $t$ with the following equations:
\begin{flalign*}
i_t &= \sigma(W_{i}x_t + U_{i}h_{t-1} + b_i) &\\
f_t &= \sigma(W_{f}x_t + U_{f}h_{t-1} + b_f) &\\
c_t &= f_tc_{t-1} + i_t \: tanh(W_{c}x_t + U_{c}h_{t-1} + b_c) &\\
o_t &= \sigma(W_{o}x_t + U_{o}h_{t-1} + b_o)  &\\
h_t &= o_t \: tanh(c_t) 
\end{flalign*}
where $x_t$ and $h_t$ are the input vectors at time $t$. $W_{*},b_{*},U_{*}$ are the parameters of the LSTM network and $*=\{o, i,f,u,c\}$. $\sigma$ is the sigmoid function and $c_t$ is the cell state. For the sake of brevity, we omit the technical details of LSTM which can be found in many related works. The output of this layer is a sequence of hidden vectors $\textbf{H} \in \mathbb{R}^{L \times d}$ where $L$ is the maximum sequence length and $d$ is the dimensional size of LSTM. It is also possible to stack layers of LSTMs on top of one another which form multi-layered LSTMs which we will adopt in our approach. 

\subsection{Neural Tensor Network}
The Neural Tensor Network \cite{DBLP:conf/nips/SocherCMN13,socher2013recursive} is a parameterized composition technique that learns the relationships between two vectors. The scoring function between two vectors are defined as follows:

\begin{equation}
s_t(\vec{q},\vec{a}) = u^T f(\vec{q}^{\:T} M^{[1:r]}\vec{a} + V[\vec{q}, \vec{a}] +b)
\label{eqn:ntn}
\end{equation} 
where $f$ is a non-linear function such as $tanh$ applied element wise. $M^{[1:r]} \in \mathbb{R}^{n \times n \times r}$ is a tensor (3d matrix). For each slice of the tensor $M$, each bilinear tensor product $\vec{q}^{\:T} M_r \: \vec{a}$ returns a scalar value to form a $r$ dimensional vector. The other parameters are the standard form of a neural network. We can clearly see that the NTN enables rich representational learning of embedding pairs by using a large number of parameters.

\begin{figure*}[ht]
  
  \centering
    \includegraphics[width=0.90\textwidth]{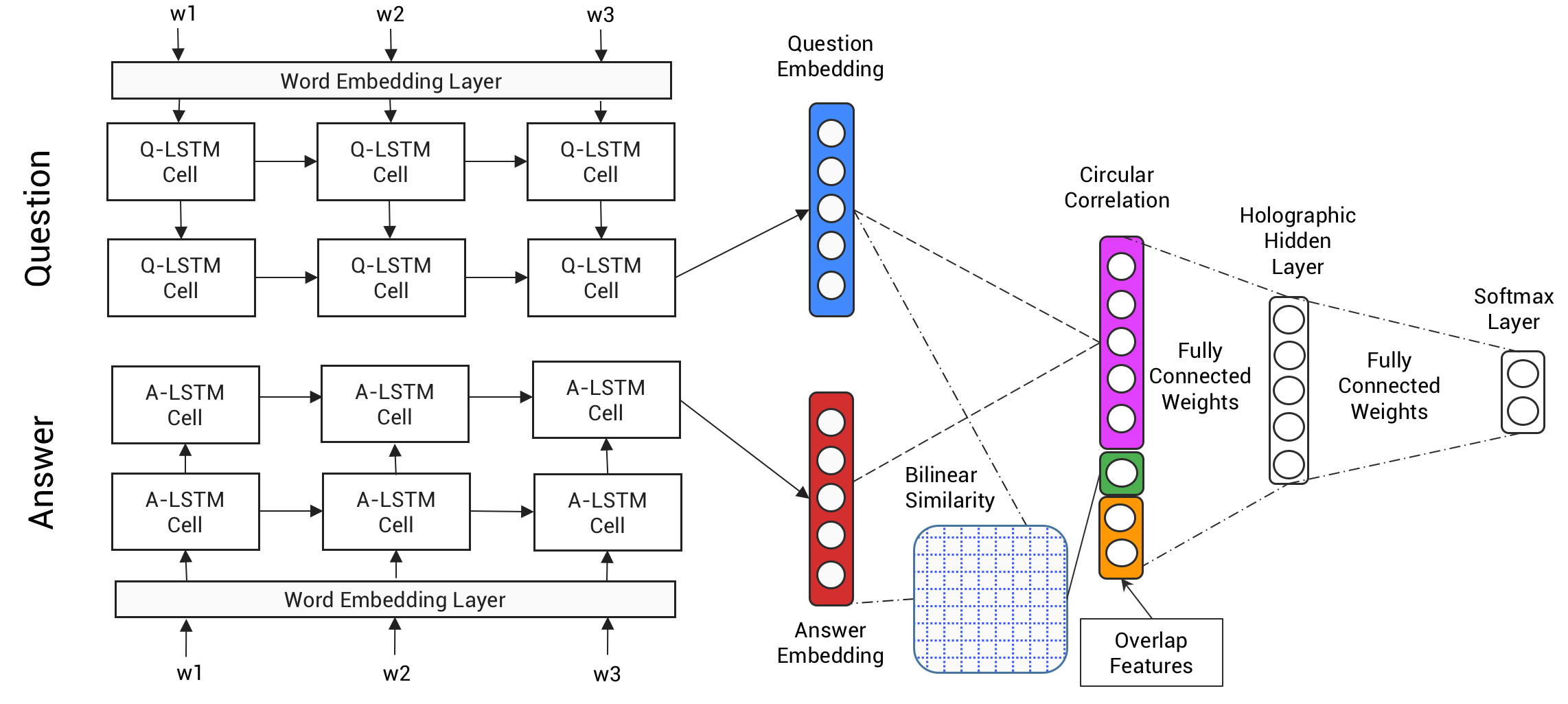}
    \caption{Holographic Dual LSTM Deep Learning Model for Ranking of QA Pairs}
    \label{fig:overall}
\end{figure*}

\section{Our Deep Learning Model}
In this section, we introduce Holographic Dual LSTM for representational learning and ranking of short text pairs. In our model, we use a pair of \textit{multi-layered} LSTMs denoted Q-LSTM and A-LSTM. First, the LSTMs learn sentence representations of question and answer pairs and subsequently holographic composition is employed to model the similarity between the outputs of Q-LSTM and A-LSTM. Finally, we pass the network through a fully connected hidden layer and perform binary classification. This is all done in an end-to-end fashion. Figure \ref{fig:overall} shows the overall architecture.  

\subsection{Learning QA Representations}
Our model accepts two sequences of indices (one for question and the other for answer) and a one-hot encoded ground truth for training. These sequence of indices are first passed through a look-up layer. At this layer, each index is converted into a low-dimensional vector representation. The parameters of this layer are $W \in \mathbb{R}^{\abs{V} \times n}$ where $V$ is the size of the vocabulary and $n$ is the dimensionality of the word embeddings. Even though these word embeddings can be learned from the training process of our model, i.e., end-to-end, we do not do so since learning word embeddings often require an extremely large corpus like Wikipedia. Therefore, we initialize $W$ with pretrained SkipGram \cite{DBLP:conf/nips/MikolovSCCD13} embeddings which is aligned with the works of \cite{DBLP:conf/sigir/SeverynM15,DBLP:journals/corr/YuHBP14}. Next, these embeddings from question and answer sequences are fed into Q-LSTM and A-LSTM respectively. Subsequently, the last hidden output from Q-LSTM and A-LSTM are taken to be the final representation for question and answer respectively.

\subsection{Holographic Matching of QA pairs}
\begin{figure}[ht]
  
  \centering
    \includegraphics[totalheight=3.2cm,width=0.28\textwidth]{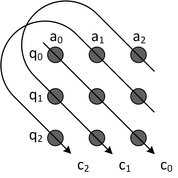}
    \caption{Circular Correlation as Compressed Tensor Product, Circular Arrows denote Summation Process, Adapted from Plate (1995) \cite{DBLP:journals/tnn/Plate95}.}
    \label{fig:ccor}
\end{figure}

The QA embeddings learned from LSTM are then passed into what we call the holographic layer. In this section, we introduce our novel compositional deep learning model for modeling the relationship between $q$ and $a$. We denote $q \: \circ \: a $ as a compositional operator applied to vectors $q$ and $a$. We employ the \textit{circular correlation} of vectors to learn relationships between question and answer embeddings.
\begin{equation}
q \circ a = q \star a
\end{equation}
where $\star :  \mathbb{R}^{d} \times \mathbb{R}^{d} \rightarrow \mathbb{R}^{d}$ denotes the circular correlation operator\footnote{For Holographic Composition, we use zero-indexed vectors for notational convenience.}. 
\begin{equation}
[q \star a]_k = \sum_{i=0}^{d-1} q_i\:a_{(k+i) \: \bmod\:d}
\end{equation}
Circular correlation can be computed as follows:

\begin{equation}
q \star a = \mathcal{F}^{-1} (\overline{\mathcal{F}(q)} \odot \mathcal{F}(a))
\end{equation}
where $\mathcal{F(.)}$ and $\mathcal{F}^{-1}(.)$ are the \textit{fast Fourier transform} (FFT) and \textit{inverse fast Fourier transform}. $\overline{\mathcal{F}(q)}$ denotes the complex conjugate of $\mathcal{F}(q)$. $\odot$ is element-wise (or Hadamard) product. Additionally, circular correlation can be viewed as a compressed tensor product \cite{DBLP:conf/aaai/NickelRP16}. In the tensor product $[q \otimes a]_{ij} = q_ia_j$ a separate element is used to store each pairwise multiplication or interaction between $q $ and $a$. In circular correlation, each element of the composed vector is a sum of multiplicative interactions over a fixed summation pattern between $q$ and $a$. Figure \ref{fig:ccor} describes this process where the circular arrows depict the summation process in which vector $\vec{c}$ is the result of composing $\vec{q}$ and $\vec{a}$ with circular correlation.  

 One key advantage of this composition method is that there are no increase in parameters. The fact that the composed vector remains at the same length of its constituent vectors is an extremely attractive quality of our proposed model. In the case where question and answer representations are of different dimensions, we can simply zero-pad the vectors to make them the same length. As circular correlation uses summation patterns, it is still possible to compose them without much implications. However, for this paper we consider that $\vec{q}$ and $\vec{a}$ to have the same dimensions. 

\subsection{Holographic Hidden Layer}
Subsequently, a fully connected hidden layer follows our compositional operator which forms the holographic layer. 
\begin{equation}
h_{out} = \sigma(W_h \ldotp [q \star a] + b_h)
\end{equation}
where $w_h$ and $b_h$ are parameters of the hidden layer and $\sigma$ is a non-linear activation function like \textit{tanh}. Traditionally, most models \cite{DBLP:conf/sigir/SeverynM15, DBLP:conf/nips/HuLLC14} use the composition operator of concatenation, denoted $ \oplus $ to combine the vectors of questions and answers. $\oplus : \mathbb{R}^{d_1} \times \mathbb{R}^{d_2} \rightarrow \mathbb{R}^{d_1 + d_2}$ simply appends one vector after another to form a vector of their lengths combined. Obviously, concatenation does not consider the relationship between latent features of QA embeddings. Thus, the relationship has to be learned from the parameters of the deep learning model, i.e., the subsequent hidden layer. In summary, the fully connected dense layer that maps $[q \star a]$ to $h_{out}$ forms the holographic hidden layer of our network.

\vspace{1em}

\textit{Incorporating Additional Features} Following the works of \cite{DBLP:conf/pkdd/BordesWU14,DBLP:conf/sigir/SeverynM15}, it is also possible (though optional) to incorporate additional features. First, we include an additional similarity measure in our model between QA embeddings. Namely, this similarity is known as the bilinear similarity which can be defined as:

\begin{equation}
sim(q,a) = \vec{q}^{\:T}M\vec{a}
\end{equation}
where $M \in \mathbb{R}^{n \times n}$ is a similarity matrix between vectors $q \in \mathbb{R}^n$ and $a \in \mathbb{R}^n$. The bilinear similarity is a parameterized approach where $M$ is an additional parameter of the network. The output of $sim(q,a)$ is a scalar value that is concatenated with $[q \star a]$. We experimented with concatenation at $h_{out}$ but empirically found it to perform worse. It is also possible to include other manual features. For example, in \cite{DBLP:conf/sigir/SeverynM15}, word overlap features $X_{feat}$ were included before the hidden layers at the join layer. The rationale for doing so is as follows: First, it is difficult to encode features like word overlap into deep learning architectures \cite{DBLP:conf/sigir/SeverynM15,DBLP:journals/corr/YuHBP14}. Secondly, word overlap features are relatively easy and trivial to implement and incorporate. As such, we are able to do the same with our model. Thus, when using external features, the inputs to the holographic hidden layer becomes a vector $[[q \star a], sim(q,a), X_{feat}]$.

\subsection{Softmax Layer}

The output from the holographic hidden layer is then passed into a fully connected softmax layer which introduces another two parameters $W_f$ and $b_f$.
\begin{equation}
p = softmax(W_f \ldotp h_{out} + b_f)
\end{equation}
where $\theta_k$ is the weight vector of the $k$th class and $x$ is the final vector representation of question and answer after passing through all the layers in the network.

\subsection{Training and Optimization}
Finally, we describe the optimization and training procedure of our network. Our network minimizes the cross-entropy loss function as follows:

\begin{equation}
L = - \sum^{N}_{i=1} \: [ y_i \log a_i + (1-y_i)\log(1-a_i)] + \lambda\norm{\theta}^{2}_2
\end{equation}

\noindent where $a$ is the output of the softmax layer. $\theta$ contains all the parameters of the network and $ \lambda\norm{\theta}^{2}_2$ is the L2 regularization. The parameters of the network can be updated by Stochastic Gradient Descent (SGD). In our network, we mainly employ the Adam \cite{DBLP:journals/corr/KingmaB14} optimizer.

\section{Discussion}
In this section, we discuss and highlight some of the interesting and advantageous properties of our proposed approach. 
\subsection{Complexity Analysis}

To better understand the merits of our proposed approach, we study the computational and memory complexity of our model with respect to the alternatives like the tensor layer.
\begin{table}[htbp]

   \centering
     \begin{tabular}{|c|c|c|}
     \hline
     Operator  & \#Parameters & Complexity \\
     \hline
     Tensor Product $\otimes$ &    $d^2$     &  $\mathcal{O}(d^2)$\\
     Concatenation $\oplus$ &      $2d$     & $\mathcal{O}({d})$ \\
     Circular Correlation $\star$ &      $d$     & $\mathcal{O}({d \log d})$ \\
     \hline

     \end{tabular}%
      \caption{Complexity Comparison between Compositional Operators}
   \label{tab:complexity_base}%
     
 \end{table}%

 \begin{table}[htbp]
   \centering
     \begin{tabular}{|c|c|c|c|}
     \hline
     Network  & \#Parameters & $d / h / k$ & On TREC\\
     \hline
     NTN  &    $d^{2}k + 2dk + 2k$     & 640 / 0 / 5& 2.1M\\
     Ours  &      $2dh + 4h$     &  640 / 64 / 0 & 41.2K\\
     \hline
  
     \end{tabular}%
      \caption{Memory Complexity Comparison between Tensor Layer and Holographic Layer}
   \label{tab:mem_complex}%
     
 \end{table}%

 \begin{table}[htbp]
   \centering

     \begin{tabular}{|c|c|}
     \hline
     Network  & Complexity\\
     \hline
     NTN  &         $\mathcal{O}({d^{2}k + 2dk + 2k})$ \\
     Ours  &            $\mathcal{O}({2dh + 4h + d\log d})$ \\
     \hline
  
     \end{tabular}%
      \caption{Complexity Comparison between Tensor Layer and Holographic Layer}
   \label{tab:run_complex}%
     
 \end{table}%
 First, Table \ref{tab:complexity_base} shows the parameter cost and complexity for the three different compositional operators. This assumes a simple example where $q,a \in \mathbb{R}^{d}$ and a vector of the same dimensionality as $q \circ a$ is used to map the composed vector into a scalar value. Clearly, the cost of a simple tensor layer is of quadratic scale which makes it difficult to scale with large $d$. Concatenation, on the other hand, doubles the length of composed vectors and increases the memory cost by a factor of two. Finally, we see that the parameter cost of circular correlation that we employ is only $d$. As such, there can be significantly less parameters in the subsequent layers. Finally, the computational complexity of circular correlation is also relatively low at $d\log d$. Next, we compare the complexity of our network and the NTN. To enable direct comparison, we exclude any additional features at the holographic hidden layer of our network and include the subsequent softmax layer. Finally, the overall network and similarity between $q$ and $a$ can be modeled as follows:\\
\begin{equation}
s_h(\vec{q},\vec{a}) = softmax(W_f^{T}f(W_h^{T}[\vec{q} \star \vec{a}] + b_h) + b_f)
\end{equation} 
where $W_h \in \mathbb{R}^{d \times r}$ is parameters at the holographic hidden layer following the composition operation, $b_h$ is the scalar bias at the hidden layer, $W_f \in \mathbb{R}^{r \times 2}$ converts the output at the hidden layer to a 2-class classification problem and $q, a \in \mathbb{R}^{d}$ where $d$ is the dimension size of the LSTM. $f(.)$ is a non-linear activation function. Note that since we consider use a softmax layer at the output, our final output $s(q,a)$ is a vector of 2 dimensions. Similarly, in the traditional tensor layer described in Equation (\ref{eqn:ntn}), we are able to simply adapt the vector $u$ to become a weight matrix of $u \in \mathbb{R}^{k \times 2}$ where $k$ is the number of slices of tensors.  

In Table \ref{tab:mem_complex} and Table \ref{tab:run_complex}, we compare the differences between the tensor layer and our holographic layer with respect to the number of parameters and computational complexity respectively. Note that $d$ is the dimensionality of QA embeddings, $h$ is the size of the hidden layer and $k$ is the number of tensor slices. To facilitate easier comparison, we do not include complexity from learning QA representations, computing bias and activation functions but only matrix and vector operations. From Table \ref{tab:mem_complex}, it is clear and evident that our approach does not require as much parameters as the NTN. We also report the optimal dimensions of the QA embeddings and hidden layer size on TREC QA. We see that our model only requires $41.2k$ parameters\footnote{We exclude word embedding and LSTM parameters in this comparison} as opposed to $2.1M$ parameters with the NTN. As such, we see that when optimal parameters required for sentence modeling is high, the cost on the subsequent matching layer becomes impractical. The problem of quadratic scale is also reflected in computational complexity. Thus, from a theoretical point of view, the holographic composition can be seen as a memory efficient and faster alternative to the neural tensor network layer.

\subsection{Associative Holographic Memories}
Holographic models of associative memories employ a series of convolutions and correlations to store and retrieve item pairs. This is sometimes referred to as convolution-correlation (holographic-like) memories \cite{DBLP:journals/tnn/Plate95}. At this point, it is apt to introduce circular convolution: $\ast :  \mathbb{R}^{d} \times \mathbb{R}^{d} \rightarrow \mathbb{R}^{d}$ which is closely related to circular correlation.

\begin{equation}
[q \ast a]_k = \sum_{i=0}^{d-1} q_i\:a_{(k-i) \bmod\:d}
\end{equation}   
In holographic associative memory models, association of vector pairs can be encoded via correlation and then decoded with circular convolution. The relationship between correlation and convolution is as follows:
\begin{equation}
q \star a = \tilde{q} \ast a
\end{equation}
\noindent where $\tilde{q}$ is the approximate inverse of $q$ such that $q_i = q_{(-i\mod d)}$. Typically, the encoding-decoding\footnote{Ideally, that the euclidean norm of $q$ and $a$ should be $\approx 1$. However, our preliminary experiments showed that adding a normalization layer did not improve the performance.} process is done via Hebbian learning in associative memory models. However, in our case, our model is holographic in the sense that correlation-convolution memories are learned implicitly via back-propagation. For example, let $h$ be the input to the hidden layer $h_{out}$, the gradients at $a$ can be represented as:

\begin{equation}
\frac{\partial E}{\partial a_i} = \sum_{k} \frac{\partial E}{\partial h_j} \: q_{(k-j\bmod d)}
\label{eqn:grad}
\end{equation}
The gradient at $a_i$ according to Equation (\ref{eqn:grad}) \cite{DBLP:conf/nips/Plate92} is equivalent to correlating $h$ with the approximate inverse of $q$. Recall that correlating with the inverse is equivalent to circular convolution. As such, this establishes the relation of our model to holographic memories. Since our main point is to illustrate these connections, we omit the entire back-propagation derivation due to the lack of space. Finally, we describe and summarize the overall advantages of employing a holographic layer in our deep architecture for learning to rank question answer pairs. 

\begin{itemize}
\item Unlike circular convolution, circular correlation is non-commutative, i.e., $q \star a \neq a \star q$. This is useful as most applications of text matching are non-symmetric. For example, questions to answers or queries to documents are not symmetric in nature. As such, we utilize correlation as the encoding operation and allow the network to decode via circular convolution while learning parameters. 
 \item The first index of the circular correlation composed vector $[q \star a]_{0}$ is the dot product of $q$ and $a$. This is extremely helpful since question answer matching requires a measure of similarity. 
 \item The computational complexity of FFT is $\mathcal{O}(d\:log\:d)$ which makes it an efficient composition. 
 \item Our composition does not increase the dimensionality of its constituent vectors, i.e, the composition of $\vec{q} \star \vec{a}$ preserves its dimensionality. On the other hand, concatenation doubles the parameter cost at the subsequent hidden layers. Furthermore, the relationship between question and answer embeddings have to be learned by the hidden layer. 
\item The association of two vectors, namely question and answer vectors are modeled end-to-end in the network. Via back-propagation, the network learns parameters that best explains this correlation via gradient descent. 
 \end{itemize}
Here it is good to note that the original holographic reduced representations \cite{DBLP:journals/tnn/Plate95} used convolution to encode and correlation to decode. However, this can be done vice versa as well \cite{DBLP:conf/aaai/NickelRP16}.

\section{Empirical Evaluation on TREC QA}
We evaluate our model on the TREC QA task of answer sentence selection on factoid based QA.
\subsection{Experiment Setup}
In this section, we introduce the dataset, evaluation procedure, metrics and compared baselines used in this experiment.

\subsubsection{Dataset}

In this task, we use the benchmark dataset provided by Wang et al. \cite{DBLP:conf/emnlp/WangSM07}. This dataset was collected from TREC QA tracks 8-13.  In this task of factoid QA, questions are generally factual based questions such as \textit{"What is the monetary value of the Nobel Peace Prize in 1989?""}. In this dataset, we are provided with \textbf{two} training sets, namely, TRAIN and TRAIN-ALL. TRAIN consists of QA pairs that have been manually judged and annotated. TRAIN-ALL is a automatically judged dataset of QA pairs and contains a larger number of QA pairs. TRAIN-ALL, being a larger dataset, also contains more noise. Nevertheless, both datasets enable the comparison of all models with respect to availability and volume of training samples. Additionally, we are also provided with a development set for parameter tuning. The results of both training sets, development set and testing set are reported in Table \ref{tab:dataset}. Finally, it is good to note that the maximum number of tokens for questions and answers are $11$ and $38$ respectively and the length of the vocabulary $\abs{V}=16468$. 

 \begin{table}[htbp]
 \small
   \centering
     \begin{tabular}{|c|c|c|c|}
     \hline
     Data & \# Questions & \# QA pairs & \% Correct \\
     \hline
     TRAIN-ALL & 1229  & 53417 & 12 \\
     TRAIN & 94    & 4718  & 7.4 \\
     DEV   & 82    & 1148  & 9.3 \\
     TEST  & 100   & 1517  & 18.7 \\
     \hline
     \end{tabular}%
      \caption{Dataset Statistics for TREC QA Dataset}
   \label{tab:dataset}%
 \end{table}%
\vspace{-2em}

 \subsubsection{Evaluation Procedure and Metrics}
\label{sec:metric}
Following the experimental procedure in \cite{DBLP:conf/sigir/SeverynM15}, we report the results of all models in two settings. In the first setting, we measure the representational learning ability of all deep learning models without the aid of external features. Conversely, in the second setting, we include an additional feature vector $X_{feat} \in \mathbb{R}^{4}$ containing the count of word overlaps (ordinary and \textit{idf} weighted) between question-answer pairs by considering inclusion and dis-inclusion of stop-words. Finally, the official evaluation metrics of MAP (Mean Average Precision) and MRR (Mean Reciprocal Rank) are used as our evaluation metrics. MRR is defined as $\frac{1}{\abs{q}} \sum^{\abs{Q}}_{q=1} \frac{1}{rank(q)}$ where \textit{rank(q)} is the rank of the first correct answer. MAP is defined as $\frac{1}{Q} \sum^{Q}_{q=1} AvgP(q)$. The MAP is the average precision across all queries $q_i \in Q$. For evaluation, we use the official \texttt{trec\_eval} script.

\subsubsection{Algorithms Compared} 
Aside from comparison with all published state-of-the-art methods, we also evaluate our model against other deep learning architectures. Since the deep learning models compared in \cite{DBLP:conf/sigir/SeverynM15} are based on Convolutional Neural Networks, we additionally compare our model with MV-LSTM \cite{DBLP:conf/aaai/WanLGXPC16} and a simple LSTM baseline in our experimental evaluation. The following lists the major and popular deep learning based models for direct comparison with our model. Model names with $\ast$ indicate that we implemented the model ourselves. 

\begin{itemize}

\item \textbf{CNN + Logistic Regression (Yu et al.)} This is the model introduced in \cite{DBLP:journals/corr/YuHBP14}. Representations of questions and answers are learned by a convolutional neural network. Subsequently, logistic regression over the learned features is used to score QA pairs. We report two settings, namely Unigram and Bigram which are also reported in their work. 
\item \textbf{CNN$^{\ast}$} We implemented a CNN model following the architecture and hyperparameters of \cite{DBLP:conf/sigir/SeverynM15}. This model includes a bilinear similarity feature while concatenating two CNN encoded sentence representations. Unlike Yu et al.'s model, this work ranks QA pairs using an end-to-end architecture. 
\item \textbf{CNTN$^{\ast}$} We implemented a neural tensor network layer to performing matching of question and answer representations encoded by convolutional neural networks. This is similar to \cite{DBLP:conf/ijcai/QiuH15} but adopts the CNN architecture and hyperparameters of Severyn et al. \cite{DBLP:conf/sigir/SeverynM15}. For the tensor layer, we use $k=5$ where k is the number of slices of the tensor.
\item \textbf{LSTM$^{\ast}$} We consider both single layer and multi-layered LSTMs as our baselines. These baseline models do not specially model the relationships between questions and answers. Instead, a concatenation operation is used to combine the QA embeddings in which the relationships between the two vectors are modeled by the hidden layer. 
\item \textbf{MV-LSTM$^{\ast}$ (Wan et al.)} This model, introduced in \cite{DBLP:conf/aaai/WanLGXPC16}, considers matching of multiple positional embeddings and subsequently applying max-pooling of \textit{top-k} interactions. For scalability reasons, we only consider the bilinear similarity setting for this model. We consider this sufficient for three reasons. First, it is reported in \cite{DBLP:conf/aaai/WanLGXPC16} that the performance benefits of tensor over bilinear is minimal. Second, it is extremely expensive computationally even when considering a bilinear similarity let alone the tensor similarity. Third, the comparison with this model mainly aims to investigate the effectiveness of multiple positional embeddings. 
\item \textbf{NTN-LSTM$^{\ast}$} We consider a Neural Tensor Network + LSTM architecture instead of the CNTN to enable fairer comparison with our LSTM based model. In this model, we replace the holographic layer in HD-LSTM with a NTN layer which forms the major comparison in this paper. For the tensor layer, similar to the CNTN, we use $k=5$ where $k$ is the number of slices of the tensor.
\item \textbf{HD-LSTM$^{\ast}$ (Ours)} Holographic Dual LSTMS is the model architecture introduced in this paper. In our HD-LSTM model, the QA representations are merged with Holographic Composition. 
\end{itemize}

\subsubsection{Implementation Details and Hyperparameters} 
\label{sec:hyp}
We implemented all deep learning architectures ourselves with the exception of Yu et al. \cite{DBLP:journals/corr/YuHBP14} which we directly report the results. To facilitate fair comparison, we implement the exact architecture of the CNN model from Severyn et al. \cite{DBLP:conf/sigir/SeverynM15} ourselves using the same evaluation procedure and optimizer. All hyperparameters were tuned on the development set using extensive grid search. We trained all models using the Adam \cite{DBLP:journals/corr/KingmaB14} optimizer with an initial learning rate of $10^{-5}$ for LSTM models and $10^{-2}$ for CNN models\footnote{This learning rate works best for CNN models with Adam} and minimized the same cross entropy loss in a point-wise fashion. We applied gradient clipping of $1.0$ of the norm for all LSTM models. 

With the exception of the single-layered LSTM and MV-LSTM, all LSTM-based models use a single-direction and multi-layered setting. The input sequences are all padded with zero vectors to the max length for questions and answers separately. The dimensionality of the LSTM models are tuned in multiples of 128 in the range of $[128,640]$ for \texttt{TRAIN} and amongst $\{512,1024\}$ for \texttt{TRAIN-ALL} in lieu of the larger dataset. The number of LSTM layers are tuned from a range of $2-4$ and batch size is fixed to $256$ for all LSTM based models. The hidden layer size for all LSTM models are amongst $\{32,64,128,256,512\}$. For regularization and preventing over-fitting, we apply a dropout of $d=0.5$ and set the regularization hyperparameter $\lambda=0.00001$. For MV-LSTM, we followed the configuration setting as stated in \cite{DBLP:conf/aaai/WanLGXPC16}. We used the pretrained word embeddings \cite{DBLP:conf/sigir/SeverynM15} of $50$ dimensions trained on Wikipedia and AQUAINT corpus. The word embedding layer is set to non-trainable in lieu of the small dataset. We trained all models for a maximum of $30$ epochs with early stopping, i.e., if the MAP score does not increase after $5$ epochs. We take MAP scores on the development set at every epoch and save the parameters of the network for the top three models on the development set. We report the best test score from the saved models. All experiments were conducted on a Linux machine with a single Nvidia GTX1070 GPU (8GB RAM).

\subsection{Experimental Results}
 \begin{table*}[htbp]
   \centering

     \begin{tabular}{|c|cccc|cccc|cc|}
     \hline
     
           & \multicolumn{4}{c|}{Setting 1 (raw)} & \multicolumn{4}{c|}{Setting 2 (with extra features)} & \multicolumn{2}{c|}{All} \\
           \cline{2-11}
           & \multicolumn{2}{c}{TRAIN} & \multicolumn{2}{c|}{TRAIN-ALL} & \multicolumn{2}{c}{TRAIN} & \multicolumn{2}{c|}{TRAIN-ALL} & \multicolumn{2}{c|}{Average} \\
           \hline
           
     Model & MAP   & MRR   & MAP   & MRR   & MAP   & MRR   & MAP   & MRR   & MAP   & MRR \\
    
     \hline
     CNN + LR (unigram) & 0.5387 & 0.6284 & 0.5470 & 0.6329 & 0.6889 & 0.7727 & 0.6934 & 0.7677 & 0.6170 & 0.6982 \\
     CNN + LR (bigram) & 0.5476 & 0.6437 & 0.5693 & 0.6613 & 0.7058 & 0.7846 & 0.7113 & 0.7846 & 0.6335 & 0.7186 \\

     \hline
     LSTM (1 layer) & 0.5731 & 0.6056 & 0.6204 & 0.6685 & 0.6406 & 0.7494 & 0.6782  & 0.7604 & 0.6280  & 0.6960   \\
     LSTM  & 0.6093 & 0.6821 & 0.5975 & 0.6533 & 0.7007 & 0.7777 & 0.7350 & 0.8064 & 0.6606 & 0.7299 \\
     CNN  & 0.5994 & 0.6584 &  0.6691  & 0.6880 & 0.7000 & 0.7469 & 0.7216 & 0.7899 & 0.6725 & 0.7208 \\
     CNTN  & 0.6154 & 0.6701 & 0.6580 & 0.6978 & 0.7045 & 0.7562 & 0.7278 & 0.7831 & 0.6764 & 0.7268 \\
     MV-LSTM  & 0.6307 & 0.6675 & 0.6488 & 0.6824 & 0.7327 & 0.7940 & 0.7077 & 0.7821 & 0.6800  & 0.7315 \\
     NTN-LSTM & 0.6274 & 0.6831 & 0.6340 & 0.6772 & 0.7225 & 0.7904 & 0.7364 & 0.8009 & 0.6800  & 0.7379 \\
  
      \hline
    
     HD-LSTM & \textbf{0.6404} & \textbf{0.7123} & \textbf{0.6744} & \textbf{0.7511} & \textbf{0.7520} & \textbf{0.8146} & \textbf{0.7499} & \textbf{0.8153} &
      \textbf{0.7042} & \textbf{0.7733} \\
   
     \hline
     \end{tabular}%
      \caption{Experimental Results of all Deep Learning Architectures on TREC QA Dataset. Best result is in boldface.}
   \label{tab:all_results_trec}%
 \end{table*}%
\label{sec:trec}
This section shows the experimental results on the TREC QA answer sentence selection task. Table \ref{tab:all_results_trec} shows the result of all deep learning architectures in four different configurations, i.e., different training sets (TRAIN vs TRAIN-ALL) and different feature settings (with and without additional features). Overall, we see that HD-LSTM outperforms all other deep learning models. The relative ranking of each deep learning architecture is in concert with our intuitions. Using a tensor layer for matching improves performance over their base models which is aligned with the results of \cite{DBLP:conf/ijcai/QiuH15}. However, we see that HD-LSTM outperforms NTN-LSTM by a significant margin across all datasets and settings despite being more efficient. This shows the effectiveness of holographic composition for rich representational learning of QA pairs despite having less parameters. Additionally, the average increase over the baseline multi-layered LSTM are $4\%$ and $5\%$ in terms of MAP and MRR respectively which can be considered significant. We also note that there is quite significant improvement with using multi-layered LSTMs over a single layered LSTM. The performance of MV-LSTM is competitively similar to NTN-LSTM in this task. However, it is good to note that MV-LSTM takes $\approx 30s$ per epoch at the bilinear setting as opposed to our model's $\approx 0.1s$ epoch with the same LSTM configurations and settings and on the same machine and GPU. On the other hand, we see that the baseline LSTM models perform worse {}than CNN based models whereby a single layer LSTM performs poorly and does almost as poor as CNN with logistic regression from Yu et al. \cite{DBLP:journals/corr/YuHBP14}. However, the NTN-LSTM and MV-LSTM perform better than the CNTN. It is good to note that our CNN model implementation achieves slightly worst results as compared to \cite{DBLP:conf/sigir/SeverynM15} because model parameters are saved at the batch level in their work while we evaluate at an epoch level instead. Nevertheless, the performance is quite similar. 

 Finally, Table \ref{tab:exp1t3} shows the results of all published models including non deep learning systems. Evidently, deep learning has significantly outperformed traditional methods in this task. It is also good to note that HD-LSTM outperforms \textbf{all} models (both deep learning and non-deep learning) even with the smaller \texttt{TRAIN} dataset. We find this result remarkable. 
 
\begin{table}[htbp]
   \centering

     \begin{tabular}{|l|cc|}
 
     \hline
     \multicolumn{1}{|c|}{Model} & MAP & MRR \\
   
     \hline
     Wang et al. (2007) \cite{DBLP:conf/emnlp/WangSM07} & 0.6029 & 0.6852 \\
     Heilman and Smith (2010) \cite{DBLP:conf/naacl/HeilmanS10a} & 0.6091 & 0.6917 \\
     Wang and Manning (2010) \cite{DBLP:conf/coling/WangM10a} & 0.5951 & 0.6951 \\
     Yao (2013) \cite{DBLP:conf/naacl/YaoDCC13}  & 0.6307 & 0.7477 \\
     Severyn \& Moschitti (2013) \cite{DBLP:conf/sigir/SeverynMTBR14} & 0.6781 & 0.7358 \\
     Yih et al. (2013) \cite{DBLP:conf/acl/YihCMP13} & 0.7092 & 0.7700 \\
     Yu et al. (2014) \cite{DBLP:journals/corr/YuHBP14} & 0.7113 & 0.7846 \\
     Severyn et al. (2015) \cite{DBLP:conf/sigir/SeverynM15} & 0.7459 & 0.8078 \\
     \hline
     HD-LSTM TRAIN & \textbf{0.7520}      & 0.8146 \\
     HD-LSTM TRAIN-ALL & 0.7499 & \textbf{0.8153}  \\
     \hline

     \end{tabular}%
     \caption{Performance Comparison of all Published Models on TREC QA Dataset.}
   \label{tab:exp1t3}%
 \end{table}
 \vspace{-2em}

\section{Empirical Evaluation on Community-based QA}
In this experiment, we consider the task of community-based question answering (CQA). We use the Yahoo QA Dataset\footnote{http://webscope.sandbox.yahoo.com/catalog.php?datatype=l\&did=10} for this purpose. The objectives of this experiment are two-fold. First, we provide more experimental evidence of the QA ranking capabilities of our model. Second, we test all models on the Yahoo QA dataset which can be considered as a large web-scale dataset with a diverse range of topics which additionally includes informal social language. 

\subsection{Experimental Setup}

We describe the dataset used, evaluation metrics and implementation details
\subsubsection{Dataset}
\vspace{-0.5em}
\begin{table}[htbp]
   \centering
   \setlength{\extrarowheight}{1.4pt}
        \begin{tabular}{|l|c|c|}
        \hline
           & \multicolumn{1}{c|}{\# QA Pairs} & \multicolumn{1}{c|}{\# Correct} \\
           \hline
       
     TRAIN & 253440 & 50688 \\
     DEV   & 31680 & 6336 \\
     TEST  & 31680 & 6336 \\
     \hline
  
     \end{tabular}%
     \caption{Dataset Statistics of Yahoo QA Dataset.}
   \label{tab:yahoo}%
 \end{table}%
 \vspace{-2.5em}

 The dataset we use is the Yahoo QA Dataset containing 142,627 questions and answers. We select QA pairs containing questions and best answers of length 5-50 tokens after filtering away non-alphanumeric characters. As such, we obtain $63,360$ QA pairs in the end. The total vocabulary size $\abs{V}$ of this dataset is 116,900. We construct negative samples for each question by sampling $4$ samples from the top $1000$ hits obtained via \textit{Lucene\footnote{http://lucene.apache.org/core/}} search. The overall statistics of the constructed dataset is shown in Table \ref{tab:yahoo}. In general, we can consider Yahoo to be a much larger dataset over TREC QA. Furthermore, in CQA, the questions and answers are generally of longer length.

\subsubsection{Baselines and Implementation Details}
For this experiment, our comparison against competitors are similar to the first experiment. Specifically, we compare our model with LSTM (baseline), MV-LSTM (Bilinear) and NTN-LSTM for LSTM-based deep learning models along with CNN and CNTN. In addition, we include the popular Okapi BM25 benchmark \cite{DBLP:conf/trec/RobertsonWJHG94} as an indicator of the difficulty of the test set. Note that our experimental results would be naturally different from \cite{DBLP:conf/aaai/WanLGXPC16} due to different train/test/dev splits and variations in the negative sampling process. The implementation details for LSTM based deep learning models are the same as Section \ref{sec:hyp}. However, due to scalability reasons and the requirement of processing significantly much more QA pairs, we limit the dimensions of the LSTM and hidden layer to be $50$. The number of layers of the LSTM is also set to $1$. For all models, we only consider a single direction LSTM. The other hyperparameters, including the choice of pretrained word embeddings, dropout and regularization are the same unless stated otherwise.

\subsubsection{Evaluation Metrics}
For this experiment we use the same metrics as \cite{DBLP:conf/aaai/WanLGXPC16}, namely the Precision@1 and MRR. P@1 can be defined as $\frac{1}{N} \sum^{N}_{i=1} \delta(r(A^{+})=1)$ where $\delta$ is the indicator function and $A^{+}$ is the ground truth. For the sake of brevity, we do not restate MRR as it is already defined in Section \ref{sec:metric}. Note that we only consider the ranking of the ground truth amongst all the negative samples for a given question.

\subsection{Experimental Results}

Table \ref{tab:yahoo_results} shows the results of the experiments on Yahoo QA Dataset. We show that HD-LSTM achieves state-of-the-art performance on the Yahoo QA Dataset. First, we notice that the performance of Okapi BM25 model is only marginal compared to random guessing. This signifies that the testing set is indeed a difficult one.

 \begin{table}[htbp]
   \centering
 
     \begin{tabular}{|l|c|c|}
     \hline
 
     Model & \multicolumn{1}{c}{P@1} & \multicolumn{1}{|c|}{MRR} \\
     \hline

     Random Guess &  0.2000     & 0.4570  \\
     Okapi BM-25 & 0.2250 & 0.4927 \\
     CNN   & 0.4125 & 0.6323\\
     CNTN & 0.4654 & 0.6687 \\
     LSTM  & 0.4875      &  0.6829\\
      NTN-LSTM & 0.5448      & 0.7309  \\
   
      \hline
     HD-LSTM &  \textbf{0.5569}    & \textbf{0.7347}  \\

     \hline
     \end{tabular}%
      \caption{Experimental Results on Yahoo QA Dataset.}
   \label{tab:yahoo_results}%

 \end{table}%
 \vspace{-2em}

 Unfortunately, we were not able to obtain any results with MV-LSTM due to computational restrictions. Specifically, each training instance involves $5000$ matching computations to be made. Hence, each epoch takes easily $\approx 3$ hours even with GPUs. Hence, from the perspective of practical applications, we can safely eliminate the MV-LSTM as an alternative. Once again, we see the trend that tensor layer improves results over their base models similar to the earlier evaluation on the TREC QA task. However, unlike the TREC datasets, the NTN-LSTM performs significantly better than the baseline LSTM probably due to the larger dataset. On the other hand, we also observe that the LSTM performs better compared to CNN on this dataset similar to the results reported in \cite{DBLP:conf/aaai/WanLGXPC16}. Finally, our HD-LSTM performs the best and outperforms the NTN-LSTM despite having less parameters and being more efficient as discussed in our complexity analysis section earlier. 

\section{Analysis of Hyperparameters}
In this section, we discuss some important observations in our experiments. such as hidden layer size on our model. In particular, we investigate the HD-LSTM, NTN-LSTM and the baseline LSTM. Due to the lack of space, we only report the hyperparameter tuning process of the TREC QA task specifically with respect to the MAP metric. 

\subsection{Effect of Embedding Dimension}
 \begin{figure}[ht]
  \centering
    \includegraphics[width=0.24\textwidth]{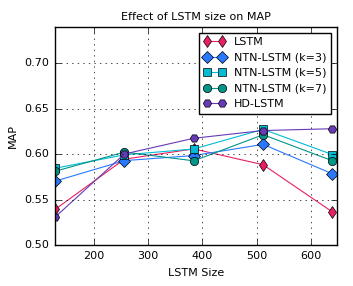}
    \caption{Effect of QA Embedding Size (LSTM Dimensions).}
    \label{fig:hp_lstm}
\end{figure}

Figure \ref{fig:hp_lstm} shows the influence of the size of the QA embedding on MAP performance on TREC \texttt{TRAIN} dataset. We investigate the NTN-LSTM at three different $k$ levels (the number of tensor slices). We see that NTN-LSTM outperforms LSTM and HD-LSTM when the dimensionality of QA embeddings are small. This is because the introduction of extra parameters of quadratic scale at the tensor layer helps the NTN-LSTM fit the data. However, when increasing the dimensions of the sentence embedding, HD-LSTM starts to steadily outperform the NTN-LSTM. Furthermore, we note that at higher LSTM dimensions, i.e., $640$, there is a steep decline in the performance of the NTN-LSTM probably due to overfitting. Overall, the performance of the baseline LSTM cannot be compared to both the HD-LSTM and NTN-LSTM. Evidently, the method used to model the relationship between the embedding of text pairs is crucial and has implications on the entire network. We see that the holographic composition allows more representational freedom in the LSTM by allowing it to have larger dimensions of text representations without possible implications.

\subsection{Effect of Hidden Layer Size}
\begin{figure}[ht]
  \centering
    \includegraphics[width=0.24\textwidth]{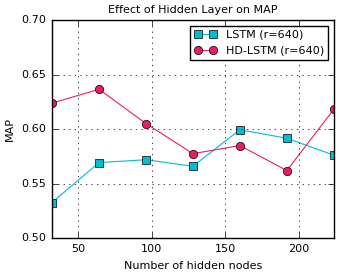}
    \caption{Effect of the Size of Hidden Layer on MAP.}
    \label{fig:hidden_layer_1}
\end{figure}

The number of nodes at the hidden layer is an important hyperparameter in our experiments due to its close proximity and direct interaction with the composition between question and answer embeddings. Note that the hidden layer size is directly related to the number of parameters of the model. In this section, we aim to study the influence of the size of the hidden layer with respect to the LSTM and HD-LSTM. Note that we are unable to directly compare with the NTN-LSTM as the tensor $M$ in the NTN layer acts like a hidden layer. Figure \ref{fig:hidden_layer_1} shows the effect of the number of hidden nodes. Evidently, we see that a smaller hidden layer size benefits the HD-LSTM. On the other hand, the performance of LSTM is only decent above a certain threshold of hidden layer size. This is in concert with our understandings of the interactions of the parameters with the composition layer. Our model requires less parameters to model the relationship between text pairs because the correlation between question and answer embeddings is modeled via holographic composition. We see that our model achieves good results even with a smaller hidden layer, i.e., 64. Contrarily, LSTM requires more parameters to model the relationship between the text embeddings. We see that HD-LSTM with a small hidden layer produces the best results.

 \section{Conclusion}
 We proposed a novel deep learning architecture based on holographic associative memories for learning to rank QA pairs. The circular correlation of vectors has attractive qualities such as memory efficiency and rich representational learning. Additionally, we overcome the problem of scaling QA representations while keeping the compositional parameters low which is prevalent in models that adopt a tensor layer. We also outperform many variants of deep learning architectures including the NTN-LSTM and CNTN in the task of learning to rank for question answering applications.

\bibliographystyle{ACM-Reference-Format}
\bibliography{references} 

\end{document}